\documentstyle[12pt]{article}

  \def\pp{{\mathchoice
          {
              \kern 1pt%
              \raise 1pt
              \vbox{\hrule width5pt height0.4pt depth0pt
                    \kern -2pt
                    \hbox{\kern 2.3pt
                          \vrule width0.4pt height6pt depth0pt
                          }
                    \kern -2pt
                    \hrule width5pt height0.4pt depth0pt}%
                    \kern 1pt
           }
            {
              \kern 1pt%
              \raise 1pt
              \vbox{\hrule width4.3pt height0.4pt depth0pt
                    \kern -1.8pt
                    \hbox{\kern 1.95pt
                          \vrule width0.4pt height5.4pt depth0pt
                          }
                    \kern -1.8pt
                    \hrule width4.3pt height0.4pt depth0pt}%
                    \kern 1pt
            }
            {
              \kern 0.5pt%
              \raise 1pt
              \vbox{\hrule width4.0pt height0.3pt depth0pt
                    \kern -1.9pt  
                    \hbox{\kern 1.85pt
                          \vrule width0.3pt height5.7pt depth0pt
                          }
                    \kern -1.9pt
                    \hrule width4.0pt height0.3pt depth0pt}%
                    \kern 0.5pt
            }
            {
              \kern 0.5pt%
              \raise 1pt
              \vbox{\hrule width3.6pt height0.3pt depth0pt
                    \kern -1.5pt
                    \hbox{\kern 1.65pt
                          \vrule width0.3pt height4.5pt depth0pt
                          }
                    \kern -1.5pt
                    \hrule width3.6pt height0.3pt depth0pt}%
                    \kern 0.5pt
            }
        }}

  \def\mm{{\mathchoice
                       {
                             \kern 1pt
               \raise 1pt    \vbox{\hrule width5pt height0.4pt depth0pt
                                  \kern 2pt
                                  \hrule width5pt height0.4pt depth0pt}
                             \kern 1pt}
                       {
                            \kern 1pt
               \raise 1pt \vbox{\hrule width4.3pt height0.4pt depth0pt
                                  \kern 1.8pt
                                  \hrule width4.3pt height0.4pt depth0pt}
                             \kern 1pt}
                       {
                            \kern 0.5pt
               \raise 1pt
                            \vbox{\hrule width4.0pt height0.3pt depth0pt
                                  \kern 1.9pt
                                  \hrule width4.0pt height0.3pt depth0pt}
                            \kern 1pt}
                       {
                           \kern 0.5pt
             \raise 1pt  \vbox{\hrule width3.6pt height0.3pt depth0pt
                                  \kern 1.5pt
                                  \hrule width3.6pt height0.3pt depth0pt}
                           \kern 0.5pt}
                       }}

\catcode`@=11
\def\un#1{\relax\ifmmode\@@underline#1\else
        $\@@underline{\hbox{#1}}$\relax\fi}
\catcode`@=12

\let\du=\du                     

\def\a{\alpha}
\def\b{\beta}
\def\c{\chi}
\def\d{\delta}

\def\f{\phi}
\def\g{\gamma}
\def\h{\eta}

\def\k{\kappa}
\def\l{\lambda}
\def\m{\mu}
\def\n{\nu}

\def\p{\pi}
\def\q{\theta}
\def\r{\rho}

\def\t{\tau}

\def\x{\xi}
\def\z{\zeta}
\def\D{\Delta}

\def\G{\Gamma}
\def\J{\Psi}
\def\L{\Lambda}
\def\O{\Omega}

\def\ve{\varepsilon}

\def\cf{{\cal F}}

\def\ck{{\cal K}}

\def\cm{{\cal M}}

\def\bo{{\raise-.5ex\hbox{\large$\Box$}}}               
\def\pa{\partial}                                       
\def\de{\nabla}                                         
\def\TH{{\raise.2ex\hbox{$\displaystyle \bigodot$}\mskip-4.7mu \llap H \;}}
\def\face{{\raise.2ex\hbox{$\displaystyle \bigodot$}\mskip-2.2mu \llap {$\ddot
        \smile$}}}                                      

\def\sp#1{{}^{#1}}                              
\def\slash#1{\rlap{\hbox{$\mskip 1 mu /$}}#1}      
\def\Tilde#1{\widetilde{#1}}                    
\def\Bar#1{\overline{#1}}                       
\def\sbar#1{\stackrel{*}{\Bar{#1}}}             

\def\abs#1{\left| #1\right|}                    
\def\leftrightarrowfill{$\mathsurround=0pt \mathord\leftarrow \mkern-6mu
        \cleaders\hbox{$\mkern-2mu \mathord- \mkern-2mu$}\hfill
        \mkern-6mu \mathord\rightarrow$}
\def\dvec#1{\vbox{\ialign{##\crcr
        \leftrightarrowfill\crcr\noalign{\kern-1pt\nointerlineskip}
        $\hfil\displaystyle{#1}\hfil$\crcr}}}           
\def\dt#1{{\buildrel {\hbox{\LARGE .}} \over {#1}}}     

\def\frac#1#2{{\textstyle{#1\over\vphantom2\smash{\raise.20ex
        \hbox{$\scriptstyle{#2}$}}}}}                   
\def\sfrac#1#2{{\vphantom1\smash{\lower.5ex\hbox{\small$#1$}}\over
        \vphantom1\smash{\raise.4ex\hbox{\small$#2$}}}} 
\def\bfrac#1#2{{\vphantom1\smash{\lower.5ex\hbox{$#1$}}\over
        \vphantom1\smash{\raise.3ex\hbox{$#2$}}}}       
\def\afrac#1#2{{\vphantom1\smash{\lower.5ex\hbox{$#1$}}\over#2}}    

\def\[{\lfloor{\hskip 0.35pt}\!\!\!\lceil}
\def\]{\rfloor{\hskip 0.35pt}\!\!\!\rceil}
\def\Lag{{\cal L}}
\def\du#1#2{_{#1}{}^{#2}}
\def\ud#1#2{^{#1}{}_{#2}}

\def\fracm#1#2{\hbox{\large{${\frac{{#1}}{{#2}}}$}}}

\def\un{\underline}
\def\fracmm#1#2{{{#1}\over{#2}}}

\def\low#1{{\raise -3pt\hbox{${\hskip 0.75pt}\!_{#1}$}}}

\def\Dot#1{\buildrel{_{_{\hskip 0.01in}\bullet}}\over{#1}}
\def\dt#1{\Dot{#1}}

\def\Tilde#1{{\widetilde{#1}}\hskip 0.015in}

\newskip\humongous \humongous=0pt plus 1000pt minus 1000pt
\def\caja{\mathsurround=0pt}
\def\eqalign#1{\,\vcenter{\openup2\jot \caja
        \ialign{\strut \hfil$\displaystyle{##}$&$
        \displaystyle{{}##}$\hfil\crcr#1\crcr}}\,}
\newif\ifdtup

\newcommand{\youngL}{
\begin{picture}(70,40)
\linethickness{\unitlength}
\multiput(0,10)(10,0){2}{\framebox(9,9){}}
\put(20,10){\framebox(29,9){$\cdots$}}
\multiput(50,10)(10,0){2}{\framebox(9,9){}}
\end{picture}}

\newcommand{\youngF}{
\begin{picture}(70,40)
\linethickness{\unitlength}
\multiput(0,10)(10,0){2}{\framebox(9,9){}}
\put(20,10){\framebox(29,9){$\cdots$}}
\multiput(50,10)(10,0){2}{\framebox(9,9){}}
\put(0,0){\framebox(9,9){$\cdot$}}
\end{picture}}

\newcommand{\youngA}{
\begin{picture}(70,40)
\linethickness{\unitlength}
\multiput(0,10)(10,0){2}{\framebox(9,9){}}
\put(20,10){\framebox(29,9){$\cdots$}}
\multiput(50,10)(10,0){2}{\framebox(9,9){}}
\multiput(0,0)(10,0){2}{\framebox(9,9){$\cdot$}}
\end{picture}}

\newcommand{\youngV}{
\begin{picture}(70,40)
\linethickness{\unitlength}
\multiput(0,10)(10,0){2}{\framebox(9,9){}}
\put(20,10){\framebox(29,9){$\cdots$}}
\multiput(50,10)(10,0){2}{\framebox(9,9){}}
\put(0,0){\framebox(9,9){$\cdot$}}
\put(10,0){\framebox(9,9){$*$}}
\end{picture}}

\newcommand{\youngH}{
\begin{picture}(70,40)
\linethickness{\unitlength}
\multiput(0,10)(10,0){2}{\framebox(9,9){}}
\put(20,10){\framebox(29,9){$\cdots$}}
\multiput(50,10)(10,0){2}{\framebox(9,9){}}
\put(0,0){\framebox(9,9){$\cdot$}}
\put(10,0){\framebox(9,9){$*$}}
\put(20,0){\framebox(9,9){$*$}}
\end{picture}}

\newcommand{\youngD}{
\begin{picture}(70,40)
\linethickness{\unitlength}
\multiput(0,10)(10,0){2}{\framebox(9,9){}}
\put(20,10){\framebox(29,9){$\cdots$}}
\multiput(50,10)(10,0){2}{\framebox(9,9){}}
\put(0,0){\framebox(9,9){$\cdot$}}
\put(10,0){\framebox(9,9){$\cdot$}}
\put(20,0){\framebox(9,9){$*$}}
\put(30,0){\framebox(9,9){$*$}}
\end{picture}}

\def\ref#1{$\sp{#1)}$}

\def\pl#1#2#3{Phys.~Lett.~{\bf {#1}B} (19{#2}) #3}
\def\np#1#2#3{Nucl.~Phys.~{\bf B{#1}} (19{#2}) #3}

\def\cqg#1#2#3{Class.~and Quantum Grav.~{\bf {#1}} (19{#2}) #3}
\def\cmp#1#2#3{Commun.~Math.~Phys.~{\bf {#1}} (19{#2}) #3}

\parskip=\medskipamount                 
\thicklines                         

\begin{document}


\thispagestyle{empty}               

\def\border{                                            
        \setlength{\unitlength}{1mm}
        \newcount\xco
        \newcount\yco
        \xco=-24
        \yco=12

        \par\vskip-8mm}

\def\headpic{                                           
        \indent
        \setlength{\unitlength}{.8mm}
        \thinlines
        \par
        \par\vskip-6.5mm
        \thicklines}

\border\headpic {\hbox to\hsize{
\vbox{\noindent NBI--HE--00--04  \hfill hep-th/0001109 \\
ITP--UH--01/00 \hfill    revised version\\
DESY 00 -- 009  \hfill    March 2000}}}

\noindent
\vskip1.3cm
\begin{center}

{\Large\bf Superconformal Hypermultiplets in Superspace}

\vglue.3in

Sergei V. Ketov \footnote{
Also at High Current Electronics Institute of the Russian Academy of Sciences,
Siberian Branch, \newline ${~~~~~}$ Akademichesky~4, Tomsk 634055, Russia}

{\it Niels Bohr Institute }\\
{\it University of Copenhagen, 2100 Copenhagen $\slash{O}$, Denmark}\\

and

{\it Institut f\"ur Theoretische Physik, Universit\"at Hannover}\\
{\it Appelstra\ss{}e 2, Hannover 30167, Germany}\\
{\sl ketov@itp.uni-hannover.de}
\end{center}
\vglue.2in
\begin{center}
{\Large\bf Abstract}
\end{center}

\noindent

We use the manifestly N=2 supersymmetric, off-shell, harmonic (or twistor) 
superspace approach to solve the constraints implied by four-dimensional N=2 
superconformal symmetry on the N=2 non-linear sigma-model target space, known 
as the special hyper-K\"ahler geometry. Our general solution is formulated in 
terms of a homogeneous (of degree two) function of unconstrained (analytic) 
Fayet-Sohnius hypermultiplet superfields. We also derive the improved 
(N=2 superconformal) actions for the off-shell (constrained) N=2 projective 
hypermultiplets, and relate them (via non-conformal deformations) to the 
asymptotically locally-flat (ALF) $A_k$ and $D_k$ series of the gravitational 
instantons. The same metrics describe Kaluza-Klein monopoles in M-theory, while
they also arise in the quantum moduli spaces of N=4 supersymmetric gauge field 
theories with $SU(2)$ gauge group and matter hypermultiplets in three spacetime
dimensions. We comment on rotational isometries versus translational 
isometries in the context of N=2 NLSM in terms of projective hypermultiplets.

\newpage

\section{Introduction}

Better understanding of non-conformal supersymmetric field theories is 
often facilitated by the study of superconformal field theories. One may, 
therefore, expect, that studying the N=2 (rigid) superconformal 
{\it Non-Linear Sigma-Models} (NLSM) in 3+1 spacetime dimensions may shed 
light on the structure of the hypermultiplet low-energy effective actions 
in quantized N=2 supersymmetric gauge field theories, as well as provide more 
insights into the moduli spaces of hypermultiplets in the type-II superstring 
compactifications on Calabi-Yau threefolds in the limit where the supergravity
decouples, since they are all governed by hyper-K\"ahler metrics. The N=2 
superconformal hypermultiplets also appear in describing the D3-brane 
world-volume field theories \cite{d3}, and in relation to the AdS/CFT 
correspondence \cite{ads}. The universal formulation of hypermultiplets is 
indispensable for those purposes. 

The most natural formulation of supersymmetry is provided by superspace. 
However, as regards N=2 supersymmetry in 1+3 dimensions, the standard N=2 
superspace is not suitable to describe off-shell hypermultiplets. Since N=2
supersymmetry in the four-dimensional NLSM amounts to hyper-K\"ahler geometry 
in the NLSM target space \cite{af}, there should be a good reason for this 
failure inside the hyper-K\"ahler geometry. As is well known, the 
hyper-K\"ahler geometry is characterized by the existence of three, 
covariantly constant, complex structures, $I^{(a)}$, $a=1,2,3$, satisfying 
a quaternionic algebra. In the conventional superspace approach one picks up a
 single complex structure to be manifest. In fact, any linear combination, 
$c_1I^{(1)}+c_2I^{(2)}+c_3I^{(3)}$, of the complex structures with some real 
coefficients $c_a$ is again a covariantly constant complex structure provided 
that $c_1^2+c^2_2+c^2_3=1$. There is, therefore, the whole sphere $S^2$ of 
the hidden complex structures on the top of the manifest one. To treat all the 
complex structures on equal footing, one should add $S^2$ to the 
hyper-K\"ahler manifold, which essentially amounts to its twistor extension. 
In the context of N=2 supersymmetry, the extension of the standard N=2 
superspace by the two-sphere $S^2$ gives rise to the off-shell 
(model-independent) formulation of a hypermultiplet with manifest (linearly
realized) N=2 supersymmetry. The twistor extension of N=2 superspace comes 
in two versions known as {\it Projective Superspace} (PSS) and 
{\it Harmonic Superspace} (HSS). The PSS appears after adding a single 
projective holomorphic coordinate of $CP^1\sim S^2$. The PSS construction 
naturally leads to {\it holomorphic} potentials for hyper-K\"ahler metrics
 via the so-called generalized Legendre transform in terms of projective 
$O(2p)$ hypermultiplets \cite{hklr,oldk}. In the HSS construction \cite{gikos}
one uses another isomorphism $S^2\sim SU(2)/U(1)$, one adds harmonics 
belonging to the group $SU(2)$, and one considers only equivariant 
(Grassmann-)analytic 
functions with respect to the $U(1)$ charge ({\it cf.} the notion of a flag 
manifold). The HSS thus naturally leads to {\it analytic} potentials for 
hyper-K\"ahler metrics. An off-shell, manifestly N=2 supersymmetric 
formulation of the most basic {\it Fayet-Sohnius} (FS) hypermultiplet is only 
possible with the infinite number of the auxiliary fields. This problem is 
elegantly solved in HSS that provides the universal formulation of a 
hypermultiplet in terms of a single unconstrained N=2 superfield in the 
analytic subspace of HSS \cite{gikos}. At the same time, the infinite number of
the auxiliary fields leads to a quite obscure connection (`bridge') between the
 HSS superfields and their physical components, which highly complicates
 a derivation of hyper-K\"ahler metrics out of the N=2 NLSM in terms of the FS
hypermultiplet superfields. The projective hypermultiplets with the finite 
numbers of the auxiliary fields, are more suitable for describing the N=2 NLSM 
metrics with isometries, either translational or rotational.

The combined constraints implied by hyper-K\"ahler geometry and conformal 
invariance on the target space geometry of the (1+3)-dimensional N=2 
supersymmetric NLSM were recently investigated by de Wit, Kleijn and Vandoren
 \cite{wkv} who called the N=2 superconformal NLSM geometry {\it special\/} 
hyper-K\"ahler. They also found remarkable relations between the
special hyper-K\"ahler, quaternionic and 3-Sasakian metrics, by using the 
{\it component} approach with on-shell N=2 supersymmetry for hypermultiplets. 
The approach adopted in ref.~\cite{wkv} is invariant under reparametrizations,
but it formulates the special hyper-K\"ahler geometry in terms of complicated 
geometrical constraints that are unavoidable in any component approach. We
use the manifestly N=2 supersymmetric HSS approoach to solve those 
constraints in terms of an unconstrained homogeneous (of degree two) function
of hypermultiplets, similarly to the standard N=2 superconformally invariant
action of N=2 vector multiplets. 

Though our general HSS construction uses the infinite number of auxiliary 
fields, we also derive the non-trivial N=2 superconformal NLSM in terms of the 
projective hypermultiplets with the finite numbers of the auxiliary fields.
As an application, we easily reproduce $A_k$ and $D_k$ series of 
four-dimensional gravitational instanton metrics by combining 
the (improved) special hyper-K\"ahler potentials with different moduli and 
adding simple non-conformal deformations to them. Those metrics naturally 
arise in modern gauge and string theories that supply more motivation for the
alternative and more transparent derivation of the metrics from HSS.

Distant physical problems in field theory often share common hyper-K\"ahler
moduli space. For example, the vacuum structure of the {\it quantized} 
$SU(n)$-based N=4 supersymmetric (pure) gauge field theory in 2+1 dimensions 
(in the Coulomb branch) is known to be equivalent to the moduli space of
$n$ BPS monopoles in the {\it classical} $SU(2)$-based (N=0) Yang-Mills-Higgs 
system in 3+1 dimensions \cite{chw}. The four-dimensional hyper-K\"ahler
spaces are necessarily self-dual, while the latter naturally arise as 
gravitational instantons in quantum gravity, or as the moduli spaces of 
solutions to the (integrable) system of Nahm equations with appropriate 
boundary conditions \cite{nahm}. 

M-theory provides the unifying framework for a study of those remarkable 
correspondences, while it also offers their explanation via dualities. 
For example, the M-theory compactification on the $A_{k-1}$ 
{\it Asymptotically Locally Flat} (ALF) self-dual space is equivalent to the 
background of $k$ parallel D6-branes in the type-IIA string picture 
\cite{sen}.~\footnote{In the case of $D_k$ ALF space, one gets $k$ D6-branes 
parallel to an orientifold $O6^-$ \cite{sen}.} Probing this background with
a parallel D2-brane gives rise to the (2+1)-dimensional N=4 supersymmetric 
effective gauge field theory with $k$ matter hypermultiplets in the D2-brane 
world-volume, whose moduli space is given by the ALF space. In the (dual) 
type-IIB string picture one gets a BPS configuration of intersecting 
D5- and D3-branes, which gives rise (in the infra-red limit) to the same 
effective field theory in the D3-brane world-volume, and whose moduli space is
given by the charge-two (centered) BPS monopole moduli space with $k$ 
singularities \cite{kc}.

Though the brane technology appears to be very efficient in establishing the
equivalence between the apparently different moduli spaces, it does not offer
any means for a calculation of the exact moduli space metrics. The standard 
hyper-K\"ahler quotient construction of the multi-monopole moduli space 
metrics from the Nahm data is very complicated in practice, so that it 
does not allow one to establish a simple and natural way of describing the
metrics. The alternative approach is provided by twistor methods whose
essential ingredients are given by holomorphic vector bundles over the twistor
space \cite{ah}. The ALF metrics can be obtained from  the (moduli-dependent)  
holomorphic potentials in the twistor space by the generalized Legendre 
transform that was originally deduced from PSS \cite{hklr}. However, even the 
generalized Legendre transform techniques are usually limited to the 
hyper-K\"ahler metrics having isometries, whereas a generic monopole moduli 
space does not have any isometries. It is, therefore, very desirable to 
develop a more universal approach to this problem.

The ALF spaces asymptotically approach 
${\bf R}^3\times S^1/\G$, where $\G$ is a discrete subgroup of $SU(2)$. After
sending the radius of $S^1$ to infinity, one gets the {\it Asymptotically 
Locally Euclidean} (ALE) metrics that asymptotically approach $R^4/\G$. 
Kronheimer \cite{kron} found their A--D--E classification into two infinite 
($A_k$ and $D_k$) and three exceptional $E_{6,7,8}$ cases,
according to the intersection matrix of their two-cycles. Since the ALE spaces
are naturally related to the enhanced symmetries, most notably, conformal
invariance, it is natural to approach an ALF moduli space metric from its 
ALE limit, being guided by hyper-K\"ahler geometry and conformal invariance on
the top of it. In the context of N=2 NLSM in superspace, the ALE potentials 
can be constructed by `switching on' vacuum expectation values of the 
hypermultiplet scalars, whereas the associated ALF potentials are obtained by  
non-conformal deformations of them. To solve the hyper-K\"ahler constraints, 
we use an unconstrained hyper-K\"ahler 
potential in HSS. The N=2 superconformal symmetry implies extra constraints on
the special hyper-K\"ahler potentials, which can be easily solved in HSS too.

Our paper is organized as follows. In sect.~2 we review some basic properties
of the special hyper-K\"ahler geometry and its relation to N=2 superconformal 
symmetry \cite{wkv}. In sect.~3  we introduce rigid N=2 
superconformal transformations in superspace \cite{fer,giosc}. The N=2
superconformal rules for various types of the N=2 hypermultiplet superfields
are discussed in sect.~4. In sect.~5 a simple general solution to the special 
hyper-K\"ahler geometry in N=2 HSS is given. In sect.~6 we turn to a 
construction of the improved (N=2 superconformal) actions of the $O(2p)$ 
projective multiplets in HSS, and then relate them to the (ALE and ALF) 
$A_k$ and $D_k$ series of gravitational instantons. Sect.~7 is devoted to our 
conclusion. All efforts were made to make our presentation as simple as 
possible. 

\section{Special hyper-K\"ahler NLSM geometry}

Let $(P_{\m},M_{\l\r};D,K_{\n})$ be the generators of the standard conformal
extension of the Poincar\'e algebra in 3+1  spacetime dimensions, 
$\m,\n,\ldots,=0,1,2,3$, where $P_{\m}$ stand for translations, 
$M_{\l\r}$ for Lorentz rotations, $D$ for dilatations and $K_{\n}$ 
for special conformal transformations. The commutation relations of the 
conformal algebra are given by a contracted $so(4,2)$ algebra. We use middle 
Greek letters to denote spacetime vector indices, whereas early Greek letters 
are reserved for the 2-component spinor indices, $\a,\b,\ldots =1,2$. A vector
 index $(\m)$ is equivalent to a bi-spinor index $(\a\dt{\a})$.

The N=2 superconformal algebra extends the conformal algebra to a contracted 
$su(2,2|2)$ superalgebra. The new generators are given by bosonic charges of 
the  $SU(2)_{\rm conf.}\times U(1)_{\rm ch.}$ internal symmetry, eight
fermionic supersymmetry charges, $Q^i_{\a}$ and $\bar{Q}^{\dt{\a}}_i$, and 
eight fermionic special supersymmetry charges, $S^i_{\a}$ and 
$\bar{S}^{\dt{\a}}_i$, where $i=1,2$. 

In the context of N=2 supersymmetric NLSM in 3+1 dimensions, N=2 supersymmetry
amounts to the hyper-K\"ahler NLSM target space $\cm$ of real dimension $4k$, 
$k=1,2,\ldots$, whose holonomy group is in $Sp(k)$ \cite{af}. 
Given the full N=2  superconformal invariance, its internal 
$su(2)_{\rm conf.}$ part implies an $su(2)$ isometry of the hyper-K\"ahler 
NLSM metric $g_{mn}$, i.e. the existence of three Killing vectors $K^m_{(A)}$
obeying Killing equations,
$$ K_{(A)}^{(m;n)}=0~,\qquad m=1,2,\ldots,4k~,\quad A=1,2,3~,\eqno(2.1)$$
and forming an $su(2)$ algebra. This non-abelian isometry is necessarily 
{\it  rotational}, i.e. it rotates complex structures in the hyper-K\"ahler 
NLSM target space $\cm$ \cite{wkv}. The dilatational invariance of a 
Riemannian manifold $\cm$ is equivalent to the existence of another (Eulerian)
 vector $X^{m}$ satisfying an equation \cite{wkv}
$$  X^{m;n}=g^{mn}~.\eqno(2.2)$$
Its geometrical significance was clarified in ref.~\cite{gr},
where eq.~(2.2) was shown to be equivalent to the following form of the metric:
$$ g_{mn}dx^mdx^n = dr^2 + r^2h_{ab}dx^adx^b~,\eqno(2.3)$$
where 
$$ x^n=(r,x^a)~,\quad a,b,c=1,2,\ldots, 4k-1~, \quad {\rm and}\quad 
h_{ab}=h_{ab}(x^c)~.\eqno(2.4)$$
Eq.~(2.3) means that $\cm$ can be considered as a cone $C(B)$ over a 
base manifold $B$ of dimension $4k-1$ \cite{gr}. The vector $X^m$ in the 
coordinates (2.4) reads
$$ X=r\fracmm{\pa}{\pa r}~,\eqno(2.5)$$
so that it is associated with the dilatations $(r,x^a)\to(\l r,x^a)$ indeed. 

Eq.~(2.2) is obvioulsy equivalent to the conformal Killing equation,
$$\Lag_X g_{mn}=X_{m;n}+X_{n;m}=2g_{mn}~,\eqno(2.6)$$
together with a condition $X_{[m;n]}=0$ or, equivalently,
$$ X_m=\pa_m f~.\eqno(2.7)$$ 
Eqs.~(2.6) and (2.7) imply that the metric $g_{mn}$ admits a potential 
$f(x^n)$,
$$ g_{mn}=\de_m\de_nf~,\quad {\rm or}\quad g_{mn}X^mX^n=2f~.\eqno(2.8)$$

In the context of hyper-K\"ahler geometry, the potential $f$ also generates
the complex structures by differentiation, so that the function $f$ is 
sometimes called the hyper-K\"ahler potential \cite{wkv,kw}. It is worth
noticing here that the potential $f$ is a constrained function in any geometry
(e.g., $\de_m\de_n\de_pf=0$), whereas we are going to introduce the 
hyper-K\"ahler (pre-)potential as an unconstrained function (sect.~5).

Though the Euler vector $X$ is not a Killing vector, it is easy to verify 
that it implies the existence of a Killing vector $Y$ in the 
presence of a covariantly constant complex structure $I$ on $\cm$ 
\cite{wkv,gr},
$$ Y^n=I\ud{n}{m}X^m~,\quad \Lag_XI=0~,\quad \[X,Y\]=0~.\eqno(2.9)$$
The second equation means that the vector $X$ is holomorphic, i.e. it preserves
the complex structure. The corresponding base manifold $B$ then carries the
so-called {\it Sasakian} structure to be obtained by projection of 
the $(I,X,Y)$ structure of $\cm$ on $B$ \cite{blair}. Since in our case 
$\cm$ is a (special) hyper-K\"ahler manifold, the base manifold $B$ 
should admit a  
{\it {\rm 3}-Sasakian} structure because of the existence of {\it three} 
independent and covariantly constant complex structures on  $\cm$,
$$ Y^n_{(A)}=I_{(A)}\ud{n}{m}X^m~,\quad \[X,Y_A\]=0~.\eqno(2.10)$$
One easily finds that the vector $X$ is tri-holomorphic, and \cite{gr}
$$ \Lag_{Y_A}I_B=-2\ve_{ABC}I_C~,\quad \[ Y_A,Y_B\]=-2\ve_{ABC}Y_C~. 
\eqno(2.11)$$
This means that the Killing vectors $Y_A$ are rotational and form an  $su(2)$ 
algebra indeed \cite{wkv}. The complex structures are invariant under 
dilatations.

The base manifold $B$ associated with a special hyper-K\"ahler manifold 
$\cm$ is called {\it {\rm 3}-Sasakian} (see ref.~\cite{bg} for a recent 
mathematical account of the 3-Sasakian manifolds). A 3-Sasakian manifold of 
real dimension $(4k-1)$ is an Einstein space, 
$$R_{ab}=(4k-1)h_{ab}~,\eqno(2.12)$$
while it takes the form of an $Sp(1)$ fibration over a quaternionic space 
\cite{bg}. In ref.~\cite{wkv} the special hyper-K\"ahler manifolds (of real 
dimension $4k$) were described as the local products of flat 4-dimensional 
space with a $(4k-4)$-dimensional quaternionic manifold, i.e. as the manifolds
 of $Sp(k-1)$ holonomy. The special hyper-K\"ahler manifolds can be equally 
defined as cones over 3-Sasakian manifolds \cite{gr}. Some applications of
the 3-Sasakian manifolds in M-theory were discussed in refs.~\cite{kw,afq}.

Our purpose is to solve the constraints implied by the special hyper-K\"ahler
geometry, in terms of an unconstrained special hyper-K\"ahler (pre-)potential. 
By using the established relation between the special hyper-K\"ahler geometry 
and the N=2 superconformal symmetry, the solution amounts to a formulation of 
the most general N=2 superconformally invariant NLSM. To do this in 
superspace, we need a HSS realization of N=2 superconformal transformations.

\section{N=2 superconformal transformations in HSS} 

The N=2 superconformal transformations in the ordinary N-extended superspaces
are known for a long time \cite{fer}. It is, therefore, straightforward to 
rewrite them to the case of N=2 supersymmetry in HSS \cite{giosc}. We follow 
ref.~\cite{giosc} in this section.

In the HSS approach the standard N=2 superspace coordinates
$Z=(x^{\m},\q^{\a}_i,\bar{\q}^i_{\dt{\a}})$ are extended by bosonic harmonics 
(or twistors) $u^{\pm i}$, $i=1,2$, belonging to the group $SU(2)$ and 
satisfying the unimodularity condition
$$ u^{+i}u^-_i=1~,\quad \Bar{u^{i+}}=u^-_i~.\eqno(3.1)$$
The original motivation for an introduction of harmonics \cite{gikos} 
was the desire to make manifest the hidden analyticity structure of the N=2 
superspace constraints defining both N=2 vector multiplets and FS 
hypermultiplets, and find their manifestly N=2 supersymmetric solutions in 
terms of unconstrained N=2 superfields.

Instead of using an explicit parametrization of the sphere $S^2$, in HSS one 
deals with the equivariant functions of harmonics, which have definite $U(1)$ 
charges defined by $U(u^{\pm}_i)=\pm 1$. The simple harmonic integration rules,
$$ \int du =1 \quad{\rm and}\quad \int du\, u^{+(i_1}\cdots
u^{+i_m}u^{-j_1} \cdots
u^{-j_n)}=0 ~~{\rm otherwise}~,\eqno(3.2)$$
are similar to the (Berezin) integration rules in superspace. In particular, 
any harmonic integral over a $U(1)$-charged quantity vanishes. The harmonic 
covariant derivatives, preserving the defining equations (3.1) in the original
(central) basis, are given by
$$ D^{++}=u^{+i}\fracmm{\pa}{\pa u^{-i}}\equiv\pa^{++}~,\quad
D^{--}=u^{-i}\fracmm{\pa}{\pa u^{+i}}~,\quad
D^{0}=u^{+i}\fracmm{\pa}{\pa u^{+i}}-u^{-i}\fracmm{\pa}{\pa u^{-i}}~~.
\eqno(3.3)$$
They satisfy an $su(2)$ algebra and commute with the standard (flat) N=2 
superspace covariant derivatives $D^i_{\a}$ and $\bar{D}_i^{\dt{\a}}\,$. The 
operator $D^0$ measures $U(1)$ charges.

The key feature of HSS is the existence of an {\it analytic} subspace 
parametrized by $$ (\z;u)=\left\{ \begin{array}{c}
x^{\a\dt{\a}}_{\rm analytic}=x^{\a\dt{\a}}-4i\q^{i\a}\bar{\q}^{\dt{\a}j}
u^+_{(i}u^-_{j)}~,~~
\q^+_{\a}=\q^i_{\a}u^+_i~,~~ \bar{\q}^+_{\dt{\a}}=\bar{\q}^i_{\dt{\a}}u^+_i~;~~
u^{\pm}_i \end{array} \right\}~,\eqno(3.4)$$
which is invariant under N=2 supersymmetry \cite{gikos}:
$$ \d x^{\a\dt{\a}}_{\rm analytic}=-4i\left( \ve^{i\a}\bar{\q}^{\dt{\a}+}
+\q^{\a+}\bar{\ve}^{\dt{\a}i}\right)u^-_i
\equiv -4i\left( \ve^{\a-}\bar{\q}^{\dt{\a}+}
+\q^{\a+}\bar{\ve}^{\dt{\a}-}\right)~,$$
$$ \d\q^+_{\a}=\ve^i_{\a}u^+_i\equiv \ve^+_{\a}~,\quad
\d\bar{\q}^+_{\dt{\a}}=\bar{\ve}^i_{\dt{\a}}u^+_i\equiv 
\bar{\ve}_{\dt{\a}}{}^+~,\quad \d u_i^{\pm}=0~,\eqno(3.5)$$
where only $\q^+_{\a,\dt{\a}}$ are present, not $\q^-_{\a,\dt{\a}}\,$~.

The usual complex conjugation does not preserve analyticity. However, it does,
 after being combined with another (star) conjugation that only acts on the
$U(1)$ indices as $(u^+_i)^*=u^-_i$ and $(u^-_i)^*=-u^+_i$. One has 
$ \sbar{u^{\pm i}}=-u^{\pm}_i$ and $\sbar{u^{\pm}_i}=u^{\pm i}$.

Analytic off-shell N=2 superfields $\f^{(q)}(\z(Z,u),u)$ of any positive 
(integral) $U(1)$ charge $q$ in HSS are defined by ({\it cf.} N=1 chiral 
superfields)  
$$D^+_{\a}\f^{(q)}=\bar{D}^+_{\dt{\a}}\f^{(q)}=0~,\quad {\rm where}\quad
D^{+}\low{\a}=D^i_{\a}u^+_i \quad {\rm and}\quad
\bar{D}^+_{\dt{\a}}=\bar{D}^i_{\dt{\a}}u^+_i~.\eqno(3.6)$$
The analytic measure reads $d\z^{(-4)}du\equiv d^4x^{\m}_{\rm analytic}
d^2\q^+d^2\bar{\q}^+du$, and it has the $U(1)$ charge $(-4)$. The
harmonic derivative $D^{++}$ in the analytic subspace (3.4) takes the form
$$D^{++}_{\rm analytic} =\pa^{++}-4i\q^{\a+}\bar{\q}^{\dt{\a}+}
\fracmm{\pa}{\pa x^{\a\dt{\a}}}~~,\eqno(3.7)$$
it preserves analyticity, and it allows one to integrate by parts. Similarly,
one easily finds that
$$D^{0}_{\rm analytic}=u^{+i}\fracmm{\pa}{\pa u^{+i}}
-u^{-i}\fracmm{\pa}{\pa u^{-i}}+\q^{\a+}\fracmm{\pa}{\pa\q^{\a+}}+
\bar{\q}^{\dt{\a}+}\fracmm{\pa}{\pa\bar{\q}^{\dt{\a}+}}~~.\eqno(3.8)$$
In what follows we omit the explicit references to the analytic subspace,
in order to simplify our notation. 

The use of harmonics allows one to make manifest (i.e. linearly realised) the
$SU(2)_R$ symmetry of N=2 supersymmetry algebra, in addition to 
manifest N=2 supersymmetry (see ref.~\cite{rec} for more details). The relation
to PSS, where the $SU(2)_R$ rotations take the form of projective
transformations, becomes clear in a particular parametrization
$$ u^+_i=(1,\x)~,\quad u^{i-}=\fracmm{-1}{1+\abs{\x}^2}\left(\begin{array}{c}
1 \\ \x \end{array}\right)~,\eqno(3.9)$$
where $\x$ is the projective (complex) $CP^1$ coordinate. 

The translational and Lorentz transformation properties of the HSS coordinates
are obvious, and we do not write them down. The transformation rules with 
respect to dilatations with the infinitesimal parameter $\r$ are also rather 
evident, being dictated by conformal weights $w$,
$$ w[x]=1~,\quad w[\q]=w[\bar{\q}]=\fracm{1}{2}~,\quad w[u]=0~.\eqno(3.10)$$
The non-trivial part of N=2 superconformal transformations is given by the 
$U(2)_{\rm conf.}$ rotations with the parameters $l^{ij}$, special conformal 
transformations with the parameters $k_{\a\dt{\a}}$, and N=2 special 
supersymmetry with the parameters $\h^i_{\a}$ and $\bar{\h}^{\dt{\a}}_i$.

The N=2 superconformal extension of the spacetime conformal transformations,
$$ \d x^{\a\dt{\a}}=\r x^{\a\dt{\a}}+k_{\b\dt{\b}}x^{\a\dt{\b}}x^{\b\dt{\a}}~,
\eqno(3.11)$$
is dictated by the requirement of preserving the unimodularity and analyticity
 conditions in eqs.~(3.1) and (3.6), respectively. As regards the non-trivial
part of the N=2 superconformal transformation laws, one finds \cite{giosc}
$$ \eqalign{
\d x^{\a\dt{\a}} & ~=~ -4i\l^{ij}u^-_iu^-_j\q^{\a+}\bar{\q}^{\dt{\a}+}
+k_{\b\dt{\b}}x^{\a\dt{\b}}x^{\b\dt{\a}} 
+4i\left( x^{\a\dt{\b}}\bar{\q}^{\dt{\a}+} \bar{\h}^-_{\dt{\b}}
-x^{\dt{\a}\b}\q^{\a+}\h^-_{\b} \right)~,\cr
\d\q^{\a+}& ~=~ \l^{ij}u^+_iu^-_j\q^{\a+} + k_{\b\dt{\b}}x^{\a\dt{\b}}\q^{\b+}
-2i (\q^{\b+}\q^+_{\b}) \h^{\a-} + x^{\a\dt{\b}}\bar{\h}^+_{\dt{\b}}~,\cr
\d\bar{\q}^{\dt{\a}+} & ~=~ -\sbar{(\d\q^{\a+})}~,\cr
\d u^+_i& ~=~ \left[ \l^{kj}u^+_ku^+_j 
+4ik_{\a\dt{\a}}\q^{\a+}\bar{\q}^{\dt{\a}+}
+4i\left( \q^{\a+}\h^+_{\a}+\bar{\h}^+_{\dt{\a}}\bar{\q}^{\dt{\a}+}\right)
\right] u^-_i~,\cr
\d u^-_i& ~=~ 0~.\cr}\eqno(3.12)$$

Since the building blocks of any invariant action in HSS are given by the 
measure, analytic superfields and HSS covariant derivatives, only their
transformation properties under rigid N=2 superconformal transformations 
are going to be relevant for our purposes. It follows from eq.~(3.12) that 
\cite{giosc}
$${\rm Ber} \fracmm{\pa(\z',u')}{\pa(\z,u)}= 1-2\L~,\quad{\rm or}
\quad \d[d\z^{(-4)}du]=-2\L[d\z^{(-4)}du]~,\eqno(3.13)$$
where the HSS superfield parameter
$$\L=-\left(\r + k_{\a\dt{\a}}x^{\a\dt{\a}}\right) +
\left(\l^{ij}+4i\q^{\a i}\h^j_{\a}+4i\bar{\h}^j_{\dt{\a}}\bar{\q}^{\dt{\a}i}
\right)u^+_iu^-_j \eqno(3.14)$$
has been introduced. Similarly, one easily finds that
$$ (D^{++})'=D^{++} - (D^{++}\L)D^0\quad {\rm and}\quad (D^0)'=D^0~.
\eqno(3.15)$$

The `truly' N=2 superconformal (infinitesimal) component parameters can thus 
be nicely encoded into the single scalar superfield $\L$ subject to the HSS 
constraint \cite{giosc}
$$ (D^{++})^2\L=0~,\eqno(3.16)$$
and the reality condition
$$ \sbar{(\L^{++})}=\L^{++}~,\quad {\rm where}\quad \L^{++}\equiv D^{++}\L~.
\eqno(3.17)$$
The transformations rules of the harmonics in eq.~(3.12),
$$ \d u^+_i=\L^{++}u^-_i~,\quad \d u^-_i=0~,\eqno(3.18)$$
together with eqs.~(3.13), (3.15), (3.16) and (3.17) represent the very 
simple and convenient description of rigid N=2 conformal supersymmetry in HSS.

\section{Superconformal hypermultiplet superfields}

The $O(2p)$ projective (or generalized tensor) multiplets in the standard N=2 
superspace are described by the N=2 superfields $L^{i_1\cdots i_{2p}}$ that 
are totally symmetric with respect to their $SU(2)_{\rm R}$ indices, being 
subject to the constraints \cite{oldk}
$$ D_{\a}^{(k}L^{i_1\cdots i_{2p})}=
\Bar{D}_{\dt{\a}}{}^{(k}L^{i_1\cdots i_{2p})}=0,\eqno(4.1)$$
and the reality condition
$$ \Bar{L}_{i_1\cdots i_{2p}}\equiv (L^{i_1\cdots i_{2p}})^*=
\ve_{i_1j_1}\cdots \ve_{i_{2p}j_{2p}}L^{j_1\cdots j_{2p}}~~.\eqno(4.2)$$
The N=2 projective multiplets are all irreducible off-shell representations 
of N=2 supersymmetry with superspin $Y=0$ and superisospin $I=p-1$ as long as
$p\geq 1$. The list of their off-shell, $8(2p-1)$ bosonic and $8(2p-1)$ 
fermionic, field components is most conveniently represented in terms of the 
$SU(2)$ Young tableaux \cite{oldk}:
$$ \begin{array}{cc}
\overbrace{\youngL}^{2p} \quad & \quad \overbrace{\youngF}^{2p} \\
L^{i_1\cdots i_{2p}} \quad & \quad  \l^{i_2\cdots i_{2p}}_{\a}  \\
 \youngA  \quad & \quad   \youngV                 \\
 N^{i_3\cdots i_{2p}} \quad & \quad V^{i_3\cdots i_{2p}}_{\a\dt{\a}} \\
 \youngH ~\quad~  & ~\quad~  \youngD \\
\c^{i_4\cdots i_{2p}}_{\a} ~\quad~  & \quad C^{i_5\cdots i_{2p}} \\ 
\end{array} \eqno(4.3)$$
where the boxes with dots and stars denote the N=2 superspace covariant
derivatives, $D^i_{\a}$ and $\bar{D}^{\dt{\a}}_i$, respectively. 
It follows from matching the numbers of the bosonic and fermionic degrees of 
freedom in eq.~(4.3) that the vector $V_{\a\dt{\a}}$ in the N=2 tensor 
multiplet $(p=1)$ is {\it conserved}, $\pa^{\a\dt{\a}}V_{\a\dt{\a}}=0$. 
The vector $V_{\a\dt{\a}}^{i_3\cdots i_{2p}}$ of any projective N=2 multiplet 
with $p>1$ is an {\it unconstrained} (general) vector field.  

The PSS naturally comes out of the efforts to construct an N=2 supersymmetric 
self-interaction of the projective multiplets in N=2 superspace 
\cite{hklr,oldk}. Let's introduce a function $G(L_A;\x,\h)$, whose arguments 
are given by some number $(A=1,2,\ldots,k)$ of the $O(2p)$ projective 
superfields (of any type) and two extra complex coordinates, $\x$ and $\h$.
Let's also impose four {\it linear} differential equations on this function, 
$$ \de_{\a}G\equiv (D^1_{\a}+\x D^2_{\a})G=0~,\quad
\D_{\dt{\a}}G\equiv (\Bar{D}^1_{\dt{\a}}+\h\Bar{D}^2_{\dt{\a}})G=0~.
\eqno(4.4)$$
It follows from the defining constraints (4.1) that a general solution to 
eq.~(4.4) reads
$$ G=G(Q_A(\x);\x)~,\quad \h=\x~,\quad 
Q_{(2p)}(\x)\equiv \x_{i_1}\cdots \x_{i_{2p}}
L^{i_1\cdots i_{2p}}~,\quad \x_i\equiv (1,\x)~,\eqno(4.5)$$
in terms of an {\it arbitrary} function $G(Q_A;\x)$. Since the function $G$ 
does not depend upon a half of the Grassmann coordinates of N=2 superspace 
by construction, its integration over the rest of the coordinates is invariant
under N=2 supersymmetry. This leads to the following universal N=2 
supersymmetric action for the projective multiplets in PSS \cite{hklr,oldk}:
$$ S[L_A] = \int d^4x\,\fracmm{1}{2\p i}\oint_C d\x\,(1+\x^2)^{-4}\Tilde{\de}^2
\Tilde{\D}^2G(Q_A,\x)+{\rm h.c.}~,
\eqno(4.6)$$
where the new derivatives, 
$$ \Tilde{\de}_{\a}\equiv \x D^1_{\a}-D^2_{\a}~,\quad \Tilde{\D}_{\dt{\a}}
\equiv \x\Bar{D}^1_{\dt{\a}}-\Bar{D}_{\dt{\a}}^2~,\eqno(4.7)$$
in the directions orthogonal to the `vanishing' directions of eq.~(4.4) have
been introduced. The integration contour $C$ in the complex $\x$-plane is 
supposed to make the action (4.6) non-trivial. The factor $(1+\x^2)^{-4}$ in 
the action (4.6) was introduced to simplify the transformation properties of 
the integrand under $SU(2)_R$ \cite{rec}. The projective variable 
$\x\in CP^1$ has the rational transformation law,
$$ \x'=\fracmm{\bar{a}\x-\bar{b}}{a+b\x}~~,\eqno(4.8)$$
whose complex parameters $(a,b)$ are constrained by the condition 
$\abs{a}^2+\abs{b}^2=1$. In general, the action (4.6) is neither conformally 
nor $SU(2)_R$ invariant.

After being expanded in components, the action (4.6) depends upon the bosonic 
Lagrange multipliers given by the vector fields $(V)$ and the scalars $(C)$, 
that can be removed via their algebraic equations of motion or by a duality
transformation. This procedure is known as the generalized Legendre transform 
\cite{hklr} that leads to a hyper-K\"ahler metric in the bosonic NLSM part of 
the theory (4.6).

The constraints (4.1) and (4.2) take the simple form in HSS,
$$ D^{++}L^{+\cdots +}=0~,\qquad \sbar{L^{+\cdots +}}=L^{+\cdots +}~,
\eqno(4.9)$$ 
where ({\it cf.} eq.~(4.5))
$$L^{+\cdots +}=u^+_{i_1}\cdots u^+_{i_{2p}}L^{i_i\cdots
i_{2p}}~.\eqno(4.10)$$ 

Requiring the invariance of the constraints (4.9) under the N=2 superconformal
transformations (3.15) and (3.18) gives rise to the covariant transformation 
laws for the projective superfields,
$$ \d L^{+\cdots +}=w\L L^{+\cdots +}~,\quad 
{\rm with}\quad w=2p~.\eqno(4.11)$$

The choice of $2p=1$ in eq.~(4.1) defines the most basic FS
hypermultiplet (with vanishing central charge), whose physical components 
comprise only scalars and spinors. It is not difficult to verify that the 
constraints (4.1) in this case imply free equations of motion, $\bo L^i=0$. 
An {\it off-shell} FS hypermultiplet is naturally described in HSS by an 
unconstrained complex analytic superfield $q^+$ of $U(1)$ charge $(+1)$
\cite{gikos}. Its free HSS action reads \cite{gikos}
$$ S[q]_{\rm free}=-\int d\z^{(-4)}du \sbar{q}{}^+D^{++}q^+~.\eqno(4.12)$$
This action is invariant under the N=2 superconformal transformations provided
that $D^{++}q^+$ transforms covariantly like $q^+$ itself, which implies 
that $q^+$ is of conformal weight one \cite{giosc},
$$ \d q^+ = \L q^+~.\eqno(4.13)$$

The free HSS equations of motion, $D^{++}q^+=0$, imply $q^+=L^i(Z)u^+_i$ 
together with the on-shell FS hypermultiplet superspace constraints,
$D_{\a}{}^{(i}L^{j)}=\bar{D}_{\dt{\a}}{}^{(i}L^{j)}=0$.

It is worth noticing that the $SU(2)_{\rm conf.}$ transformations, which are
the part of the N=2 superconformal symmetry, are {\it different} from the 
$SU(2)_{\rm R}$ transformations, with the latter being defined by their 
natural action on the Latin indices, $i,j=1,2$, as 
$\d_{SU(2)_{\rm R}}u^{i\pm}=\l\ud{i}{j}u^{j\pm}$, etc. 
For example, as regards the free off-shell theory (4.12), one finds 
\cite{giosc} 
$$ \d_{SU(2)_{\rm R}}q^+=\d_{SU(2)_{\rm conf.}}q^+ +
\l^{ij}u^-_iu^-_jD^{++}q^+~,\eqno(4.14)$$
so that the $SU(2)_{\rm conf.}$ and $SU(2)_{\rm R}$ transformations coincide 
only if $D^{++}q^+=0$, i.e. on-shell.

\section{General N=2 superconformal NLSM actions}

We are now in a position to discuss the N=2 superconformal hypermultiplet 
actions in HSS. We use the pseudo-real $Sp(1)$ notation for a single FS 
hypermultiplet superfield,
$$ q^+_a=(\sbar{q}{}^+,~q^+)~,\quad a=1,2~, \quad q^{a+}=\ve^{ab}q^+_b~,
\eqno(5.1)$$
and further generalize it to the case of several FS hypermultiplets, 
$q^{a+}\to q^{A+}$ and $q^+_A=\O_{AB}q^{B+}$, with a constant (antisymmetric) 
$Sp(k)$-invariant metric $\O_{AB}$,  $A,B=1,\ldots,2k$.

First, we recall that the most general N=2 supersymmetric NLSM can be most
naturally formulated in HSS, in terms of the FS hypermultiplet superfields, as
$$ S_{\rm NLSM}[q] =-\fracmm{1}{\k^2}
\int d\z^{(-4)}du \left[ \fracm{1}{2}q_A^{+}D^{++}q^{A+} 
+\ck^{(+4)}(q^{A+},u^{\pm}_i)\right]~,\eqno(5.2)$$ 
where the real analytic function $\ck^{(+4)}=\sbar{\ck^{(+4)}}$ of $U(1)$ 
charge $(+4)$ is called a hyper-K\"ahler (pre-)potential 
\cite{giosk}.~\footnote{We now choose our HSS 
superfields to be dimensionless, by the use of the dimensionful coupling 
\newline ${~~~~~}$ constant $\k$ in front of their actions.} By manifest N=2
supersymmetry of the NLSM action (5.2), the NLSM metric must be 
hyper-K\"ahler for any choice of $\ck^{(+4)}$. Unfortunately, an explicit
 general relation between a hyper-K\"ahler potential and the corresponding 
hyper-K\"ahler metric is not available (see, however, refs.~\cite{giosk,met}
for the explicit hyper-K\"ahler potentials of the (ALE) multi-Eguchi-Hanson, 
(ALF) multi-Taub-NUT and Atiyah-Hitchin metrics, and their derivation from
the NLSM (5.2) in terms of FS hypermultiplet superfields, in HSS).    

Eq.~(5.2) formally solves the hyper-K\"ahler constraints on the NLSM metric
in terms of an arbitrary function $\ck^{(+4)}$. It is, therefore, quite natural
to impose extra N=2 superconformal invariance on this function, in order to
determine a general solution to the special hyper-K\"ahler geometry, since 
the free part (4.12) of the action (5.2) is automatically N=2 superconformally 
invariant. In general, the invariance of the HSS action (5.1) merely implies
the invariance of the HSS Lagrangian up to a total derivative, because of the
identity
$$ \int d\z^{(-4)}du D^{++}X^{++}=0~.\eqno(5.3)$$
However, in the case of {\it unconstrained} FS analytic superfields $q^+$, 
the hyper-K\"ahler potential should be invariant too.  Eqs.~(3.13), (3.18) 
and (4.13) now imply
$$ \L\left[ \fracmm{\pa\ck^{(+4)}}{\pa q^{A+}}q^{A+} -2\ck^{(+4)}\right]
+\L^{++}\fracmm{\pa\ck^{(+4)}}{\pa u^+_i}u^-_i=0~.\eqno(5.4)$$
Equation (5.4) is equivalent to two constraints,
$$ \fracmm{\pa\ck^{(+4)}}{\pa q^{A+}}q^{A+}=2\ck^{(+4)}\quad {\rm and}\quad
 \fracmm{\pa\ck^{(+4)}}{\pa u^+_i}=0~.\eqno(5.5)$$
This means that the {\it special} hyper-K\"ahler potential of the N=2
superconformally invariant NLSM, in terms of the analytic FS superfields 
$q^{A+}$ in HSS, is given by a {\it homogeneous (of degree two)} function 
$\ck^{(+4)}(q^{A+},u^-_i)$ of $q^{A+}$. There is no restriction on the 
dependence of $\ck^{(+4)}$ upon $u^-_i$, while it should be independent upon 
$u^+_i$. This represents one of our main new results in this paper.

Our  simple description of the N=2 superconformal {\it hypermultiplet} 
actions in terms of the off-shell N=2 superfields is to be compared to the 
well-known description of the (abelian) N=2 {\it vector} multiplet actions 
in the standard N=2 (chiral) superspace \cite{fan}~,
$$ S[W]=\int d^4xd^4\q\, \cf(W_A) +{h.c.}~,\eqno(5.6)$$
in terms of the N=2 restricted chiral superfields $W_A$ representing the N=2
abelian gauge field strengths. The N=2 superconformal invariance of the action
(5.6) implies that $\cf(W_A)$ is a {\it homogeneous (of degree two)} 
function of $W_A$ \cite{van}. The special K\"ahler geometry, associated with 
the scalar NLSM part of the action (5.6), and the special hyper-K\"ahler 
geometry (sect.~2) are, however, very different at the level of components, 
as well as in the geometrical terms.

A non-trivial special hyper-K\"ahler potential exists even in the case of
a single FS hypermultiplet, e.g.,
$$ \ck^{(+4)}(\sbar{q}{}^+,q^+,u^-_i)=
C\left[\fracmm{\sbar{q}{}^+q^+}{\sbar{q}{}^+u_2^--q^+u^-_1}
\right]^2~,\eqno(5.7)$$
where $C$ is a real constant. Eq.~(5.7) is not invariant with respect to 
$SU(2)_{\rm R}$ because of its explicit dependence upon harmonics, whereas it 
is invariant under the $SU(2)_{\rm conf.}$ part of the N=2 superconformal 
symmetry by construction. The corresponding special hyper-K\"ahler metric 
interpolates between the Eguchi-Hanson (ALE) metric described by a 
hyper-K\"ahler potential 
$$ \ck^{(+4)}_{\rm EH}=\left[\fracmm{\x^{++}}{\sbar{q}{}^+u_2^--q^+u^-_1}
\right]^2~,\eqno(5.8)$$
in the limit $(\sbar{q}{}^+q^+)\to \x^{++}=\x_{ij}u^{i+}u^{j+}$ with constant 
$\x_{ij}$, and the Taub-NUT (ALF) metric described by a hyper-K\"ahler 
potential (in the limit~ $\sbar{q}{}^+u_2^--q^+u^-_1=const.$)
$$ \ck^{(+4)}_{\rm Taub-NUT}=Const.\left(\sbar{q}{}^+q^+\right)^2~.
\eqno(5.9)$$

\section{N=2 superconformal projective multiplets}

We now turn to a HSS construction of the N=2 superconformal (improved) actions
for the $O(2p)$ projective multiplets introduced in sect.~4. Unlike the FS
hypermultiplets described by unconstrained (analytic) superfields in HSS, the 
projective multiplets are described by the constrained (off-shell) analytic 
superfields that give rise to the finite numbers of the auxiliary fields. It 
is, therefore, straightforward to deduce the component hyper-K\"ahler metrics 
out of their HSS actions. The N=2 supersymmetric selfinteraction (4.6) of the 
projective multiplets ($p<\infty$) in PSS is known to be merely a subclass of 
the most general N=2 NLSM described by eq.~(5.2) in HSS \cite{ten}, while the 
projective superfields are of higher conformal weight than FS superfields --- 
see eqs.~(4.11) and (4.13). We should, therefore, expect severe constraints on
 the N=2 superconformal actions in terms of the projective multiplets. This 
is known to be the case for the standard N=2 tensor multiplet $(p=1)$ indeed, 
whose N=2 improved action was constructed many years ago, first in components 
\cite{imc}, then in N=1 superspace \cite{imp} and N=2 PSS \cite{hklr}, and 
finally in HSS \cite{imh}. We begin with a simple derivation of the improved 
N=2 tensor multiplet action in HSS, and then discuss its N=2 superconformal 
generalizations and non-conformal deformations.
 
On dimensional reasons, the most general N=2 supersymmetric action of a 
single N=2 tensor multiplet superfield $L^{++}$, subject to the constraints
(4.9), reads in HSS as \cite{ten} 
$$ S[L]=\fracmm{1}{\k^2} \int d\z^{(-4)}du \Lag^{(+4)}(L^{++};u^{\pm}_i)~.
\eqno(6.1)$$ 
A free bilinear action in $L^{++}$ is obvioulsy not N=2 superconformally 
invariant, so that it has to be {\it improved} in some non-trivial way. A
power series in terms of $L^{++}$, as the naive {\it Ansatz} for the HSS 
Lagrangian $\Lag^{(+4)}$, also does not work here, because of  the  need to
balance the conformal weights defined by eqs.~(3.13) and (4.11). A resolution 
of this problem was suggested in ref.~\cite{imc}, where it was noticed 
that the improved Lagrangian must be topologically non-trivial, i.e. it should
contain a Dirac-like string of singularities parametrized by an arbitrary  
$SU(2)$ triplet of constants $c^{ij}$, 
$$c^{ij}=c^{ji}~,\quad \Bar{(c^{ij})}=\ve_{ik}\ve_{jl}c^{kl}~,\quad
c^2=\fracm{1}{2}c_{ij}c^{ij}~.\eqno(6.2)$$
This essentially amounts to extracting a `fake' vacuum 
expectation value out of the N=2 tensor superfield $L^{ij}$,
$$ L^{ij}=c^{ij}+l^{ij}~,\quad{\rm or,~equivalently,}\quad L^{++}
=c^{++}+l^{++}~.\eqno(6.3)$$
The N=2 superconformal invariance of the {\it action} makes it to be 
independent upon the constants $c^{ij}$ because of the $SU(2)_{\rm conf.}$ 
symmetry. Since the normalization of $c^2$ can always be changed by 
dilatations, we temporarily set $c^2=1$ for simplicity of 
our calculations in what follows. The definitions 
$$ c^{++}=c^{ij}u^+_iu^+_j~,\quad  c^{+-}=c^{ij}u^+_iu^-_j~,\quad
 c^{--}=c^{ij}u^-_iu^-_j~,\eqno(6.4)$$
imply the identities \cite{imh}
$$ D^{++}c^{--}=2c^{+-}~,\quad D^{++}c^{+-}=c^{++}~,\quad 
D^{++}c^{++}=0~,\eqno(6.5)$$ 
and
$$ c^{++}c^{--}-(c^{+-})^2=c^2=1~,\eqno(6.6)$$
with the latter being the corollary of the completeness relation for 
harmonics,
$$ u^{i+}u^-_j- u^{+}_ju^{i-}=\d\ud{i}{j}~. \eqno(6.7)$$

Equations (6.2), (6.3) and (6.5) imply that $l^{++}$ also satisfies the 
initial off-shell constraints (4.9), 
$$D^{++}l^{++}=0~,\qquad \sbar{l^{++}}=l^{++}~.\eqno(6.8)$$

The natural {\it Ansatz} for the improved N=2 tensor multiplet action in HSS
is given by \cite{imh}
$$ S[L]_{\rm impr.} =\fracmm{1}{\k^2} \int d\z^{(-4)}du (l^{++})^2
f(y)~,\quad y\equiv l^{++}c^{--}~,\eqno(6.9)$$ 
where the function $f(y)$ is at our disposal. Since the action in question is
supposed to improve the naive (quadratic) one, the function $f$ should obey 
the boundary condition
$$f(0)=1~.\eqno(6.10)$$
A more general HSS {\it Ansatz} for the HSS Lagrangian may include a term 
$(c^{++})^2g(y,c)$ in eq.~(6.9), with yet another function $g(y,c)$ to be 
discussed below. 

The identities (6.5) and (6.6) together with the constraint (6.8) further 
imply that
$$ 2y(D^{++})^2y-(D^{++}y)^2=4(l^{++})^2\quad {\rm and}\quad
 (D^{++})^3y=0~.\eqno(6.11)$$

The N=2 superconformal transformation laws of the new variables $l^{++}$ and 
$y$ follow from their definitions in eqs.~(6.3) and (6.9), by the use of
eqs.~(3.18) and (4.11) with $p=1$ and $w=2$. We find
$$ \d l^{++} = 2\L l^{++} + 2D^{++}\left(\L c^{+-} - \L^{++}c^{--}\right)~,
\eqno(6.12)$$
and
$$\eqalign{
\d y & = 2\L y +2\left[ \L c^{++}-\L^{++} c^{+-}\right]c^{--} \cr
& = 2\L y +2D^{++}\left[ \L c^{--}c^{+-} -\L^{++}(c^{--})^2\right]
+4c^{+-}\left[\L^{++}c^{--}-\L c^{+-}\right]~.\cr}\eqno(6.13)$$

Varying the action (6.9) by the use of eqs.~(3.13), (6.12) and (6.13), 
integrating by parts via eq.~(5.3), and using the identities (6.11) yield 
$$ \eqalign{
\d \int d\z^{(-4)}du (l^{++})^2f(y)=& -\int d\z^{(-4)}du \L(D^{++}y)^2\left[
y(y+1)f'' +\frac{1}{2}(7y+6)f' +\frac{3}{2}f\right] \cr
& + \int d\z^{(-4)}du \L^{++}(D^{++}y)\left[ y^2f''+y(6-y)f'-yf\right]=0~.
\cr}\eqno(6.14)$$ 
Note that the second line of eq.~(6.14) is also a total derivative in the 
case of a single N=2 tensor multiplet. The N=2 superconformal invariance of 
the action (6.9) thus amounts to the second-order ordinary differential 
equation on the function $f(z)$, 
$$ z(1-z)f_{zz} +\frac{1}{2}(6-7z)f_z -\frac{3}{2}f=0~,\qquad z=-y~.
\eqno(6.15)$$
This is the very particular hyper-geometric equation, whose well-known 
general form depending upon three parameters $(\a,\b,\g)$ is given by 
$$ z(1-z)F_{zz}+\left[\g-(\a+\b+1)z\right]F_z-\a\b F=0~.\eqno(6.16)$$
Hence, a regular solution to our problem (6.15) with the boundary condition 
(6.10) is given by the hyper-geometric function $F(\a,\b,\g;z)$ with 
$\a=1$, $\b=\frac{3}{2}$ and $\g=3$. It appears to be an elementary function 
$$ f(z)=F(1,\frac{3}{2},3; z)=\left[ \fracmm{1+\sqrt{1-z}}{2}\right]^{-2}~,
\eqno(6.17)$$
in full agreement with ref.~\cite{imh}, where the recursive methods were used.
As was demonstrated in ref.~\cite{imh}, integration over the Grassmann and 
harmonic coordinates of HSS in the action defined by eqs.~(6.9) and (6.17),
$$ S[L]_{\rm impr.} = \fracmm{4}{\k^2}\int d\z^{(-4)} du\left[ 
\fracmm{L^{++}-c^{++}}{1+\sqrt{1+(L^{++}-c^{++})c^{--}/c^2}}\right]^2~~, 
\eqno(6.18)$$
results in the improved component action of ref.~\cite{imc}. The equivalent
PSS action (4.6) has a holomorphic potential \cite{hklr}
$$ G(Q(\x),\x) = (Q(\x)-c(\x))\left[ \ln (Q(\x)-c(\x))-1\right]~,\eqno(6.19)$$
where $Q=\x_i\x_jL^{ij}$ and $c(\x)=\x_i\x_jc^{ij}$ with $\x_i=(1,\x)$, while 
the contour $C$ in the complex $\x$-plane encircles the roots of a quadratic 
equation $\left.(Q-c)(\x)\right|=0$. Like the HSS action (6.9), the equivalent
 PSS action (4.6) with the potential (6.19) does not depend upon the constants
 $c^{ij}$.

It is not difficult to verify that another {\it Ansatz} for the N=2 tensor 
multiplet HSS Lagrangian of the form $(c^{++})^2g(y,c^{+-})$ gives rise to 
another N=2 superconformal invariant,
$$ \int d\z^{(-4)} du (c^{++})^2 L^{++}c^{--}~. \eqno(6.20)$$
However, it vanishes after integration over harmonics and the anticommuting
N=2 superspace coordinates, because of the conservation law 
$\pa_{\a\dt{\a}}V^{\a\dt{\a}}=0$ for the vector component of the N=2 tensor 
multiplet in eq.~(4.3).

The N=2 superconformally invariant action (6.18) leads to a {\it flat} NLSM 
metric (in disguise) after the generalized Legendre transform \cite{imc}. The 
improved N=2 tensor multiplet action can, nevertheless, serve as the key 
building block for a construction of non-trivial four-dimensional 
hyper-K\"ahler metrics. For example, the $A_k$ series of the ALE gravitational
instanton metrics arise when one sums the improved N=2 tensor multiplet 
Lagrangians with {\it different} moduli $c_a^{ij}$, 
$$ S[L]_{{\rm ALE}-A_k} = \fracmm{4}{\k^2}\int d\z^{(-4)} du \sum^{k+1}_{a=1}
 \left[
\fracmm{L^{++}-c_a^{++}}{1+\sqrt{1+(L^{++}-c_a^{++})c_a^{--}/c_a^2}}\right]^2
~~.\eqno(6.21)$$
The associated PSS potential reads
$$ G(Q(\x),\x) = \sum^{k+1}_{a=1}(Q(\x)-c_a(\x))\left[ 
\ln (Q(\x)-c_a(\x))-1\right]~,\eqno(6.22)$$
while its generalized Legendre transform is known to lead to the $A_k$ ALE
metrics indeed \cite{ir}. Another simple non-conformal deformation is given by
 the naive (bilinear) action of the N=2 tensor multiplet,
$$ S[L]_{\rm naive}= m \int d\z^{(-4)} du (L^{++})^2 ~.\eqno(6.23)$$ 
After being added to the action (6.21), it leads to the $A_k$ series of the
ALF (multi-Taub-NUT) metrics with a Taub-NUT mass parameter $m$ (see e.g.,
ref.~\cite{rec}). As is clear from eq.~(6.22), the moduli of the $A_k$ metrics
 are naturally described by fixed real sections $c_a(\x)$ of the $O(2)$ 
holomorphic bundle, while one of them can be arbitrarily chosen. In the 
context of M-theory/type-IIA superstring compactification (sect.~1), the $A_k$
 moduli describe the D6-brane positions in the transverse space.

Having established the improved action of a {\it single} N=2 tensor multiplet
 $(p=1)$, it is natural to look for N=2 superconformal actions in terms of 
{\it several\/} N=2 tensor multiplets or the {\it higher} $O(2p)$ projective 
multiplets as well (a geometrical motivation of the latter is discussed at the
 end of this section). 

A generalization of the {\it Ansatz} (6.9) to the case of several N=2 tensor 
multiplets $(p=1)$ is given by
$$ S[L_a]= \fracmm{1}{\k^2} \int d\z^{(-4)}du \sum_{a,b=1}^q l_a^{++}l_b^{++}
f_{ab}(\{y\})~,\quad a=1,2\ldots,q~,\eqno(6.24)$$ 
where $f_{ab}(\{y\})$ is the symmetric matrix of $q(q+1)/2$ functions 
depending upon $q$ variables, $\{y\}=(l_1^{++}c^{--},\ldots,l_q^{++}c^{--})$, 
with {\it the same} $c^{--}$. Requiring the action (6.24) be invariant under 
the N=2 superconformal transformations gives rise to a system of $q(q+1)/2$ 
second-order ordinary differential equations,~\footnote{We denote 
differentiations by commas, like in general relativity.}
$$ \sum^q_{a=1} \left[(q+\sum^q_{b=1}y_b)y_a+2(1+y_a)\right]f_{a(c,d)}
+\sum^q_{a=1}(1-\frac{3}{2}y_a)f_{cd,a}-\fracm{3}{2}f_{cd}=0~,\eqno(6.25)$$
and extra consistency condition on the vector
$$V_a\equiv \sum_{b,c=1}^q y_b(2-y_c)f_{ab,c}-\sum^q_{b=1}f_{ab}y_b\eqno(6.26)
$$
to be a total derivative, i.e. $\pa_aV_b-\pa_bV_a=0$. This gives rise to 
the additional equations
$$ \sum_{a,b=1}^q y_a(2-y_b)f_{a[c,d]b}-2\sum_{a=1}^qf_{a[c,d]}y_a=0 
\eqno(6.27)$$
that make the full set of $q^2$ equations to be overdetermined. We are unaware
 of any non-trivial solutions to these equations, except of the one given by a 
non-interacting sum of the improved actions for each N=2 tensor multiplet.

We now turn to a single $O(4)$ projective multiplet $L^{++++}\equiv L^{(+4)}$
satisfying the constraints (4.9), and construct its improved action in HSS. 
Instead of writing down a new {\it Ansatz}, it is much simpler to use the 
already established improved action (6.18) for the $O(2)$ projective (tensor) 
multiplet, and then take into account the known transformation property (4.11)
of $L^{(+4)}$ with $p=2$ and $w=4$. The latter implies that $L^{(+4)}$ 
transforms as $(L^{++})^2$ under the N=2 superconformal transformations. 
By using the obvious identities,
$$ L^{++}c^{--}=\sqrt{(L^{++})^2(c^{--})^2} \eqno(6.28)$$
and
$$ (L^{++}-c^{++})^2=(L^{++})^2-4(D^{++})^2
\sqrt{(L^{++})^2(c^{--})^2}+(c^{++})^2~,\eqno(6.29)$$
we can simply substitute  $(L^{++})^2$ by  $L^{(+4)}$ in eq.~(6.18). It yields
the N=2 superconformally invariant (improved) action of an $O(4)$ projective 
multiplet in the form
$$ \eqalign{
S[L^{(+4)}]_{\rm impr.} = & \fracmm{4}{\k^2}\int d\z^{(-4)} du 
\left[1+\sqrt{1+\fracmm{1}{c^2}\sqrt{L^{(+4)}(c^{--})^2}
-\fracmm{c^{++}c^{--}}{c^2}}\right]^{-2}\times \cr
& \times \left\{ L^{(+4)}
\left(1-\fracmm{8c^{++}c^{--}}{\sqrt{L^{(+4)}(c^{--})^2}}\right)
+(c^{++})^2 \right\}~. \cr} \eqno(6.30)$$
This action is one of our main new results in this paper. The associated PSS
potential is obtained from eq.~(6.19) after a substitution $Q_{(2)}\to
\sqrt{Q_{(4)}}$~, i.e.  
$$ G(Q_{(4)}(\x),\x) = \left(\sqrt{Q_{(4)}(\x)}-c(\x)\right)\left[ \ln 
\left(\sqrt{Q_{(4)}(\x)}-c(\x)\right)-1\right]~,\eqno(6.31)$$
while the integration contour $C_2$ in the complex $\x$-plane is now given 
by two circles around the branch cuts of $\sqrt{Q_{(4)}}$ \cite{ir,crw}.

It is worth noticing that the HSS superfield $\sqrt{L^{(+4)}(c^{--})^2}$
transforms covariantly under the N=2 superconformal transformations with
conformal weight $w=2$, 
$$ \d \sqrt{L^{(+4)}(c^{--})^2}=2\L\sqrt{L^{(+4)}(c^{--})^2}~.\eqno(6.32)$$ 
This implies the existence of another {\it non-trivial\/} N=2 superconformal 
invariant that originates from eq.~(6.20) and has the form
$$ S[L^{(+4)},c]_{\rm ext.} =\int d\z^{(-4)} du (c^{++})^2  
\sqrt{L^{(+4)}(c^{--})^2}~. \eqno(6.33)$$
The holomorphic PSS potential 
associated with the HSS Lagrangian (6.33) is obvious,
$$  G(Q_{(4)}(\x))_{\rm ext.}= \sqrt{Q_{(4)}(\x)}~.\eqno(6.34)$$

We are now in a position to make use of the new improved action (6.30) of a 
single $O(4)$ projective multiplet. The generalized Legendre transform in 
application to eq.~(6.31) is supposed to yield a {\it free} NLSM metric 
(in disguise), similarly to the improved $O(2)$ projective multiplet action in
eqs.~(6.18) and (6.19). However, a sum of the improved actions (6.30) and 
(6.33) with {\it different\,} moduli $c_a^{ij}$,
$$ S[L^{(+4)}]_{{\rm ALE}-D_k}=\sum_{a=1}^{k} \left(S[L^{(+4)},c_a]_{\rm impr.}
+S[L^{(+4)},-c_a]_{\rm impr.}\right)+ S[L^{(+4)},c_0]_{\rm ext.}~,\eqno(6.35)$$
gives rise (after the generalized Legendre transform) to the N=2 NLSM whose
non-trivial metric can be identified with the ALE $D_k$ metric \cite{kc,crw}. 
The $Z_2$ symmetric combination of the improved terms in eq.~(6.35) is 
necessary to produce the dihedral group $D_k$ out of the cyclic $C_k$ group. 
Similarly, the ALF series of $D_k$ metrics are obtained after adding to
eq.~(6.35) a non-conformal deformation ({\it cf.} eq.~(6.23)),
$$ m \int d\z^{(-4)}du \,L^{(+4)}~. \eqno(6.36)$$
In the context of M-theory/type-IIA compactification (sect.~1), the parameters 
$\{c^{ij}_a\}$ in the action (6.35) describe the positions of D6-branes and 
an orientifold in the transverse directions. In the context of the related 
three-dimensional N=4 supersymmetric gauge field theory with $k$ matter 
hypermultiplets in the probe D2-brane world-volume, the moduli $\{c^{ij}_a\}$ 
parametrize the quantum moduli space (= ALF $D_k$ with $k$ singularities) of 
the low-energy effective field theory, being related to the positions of 
monopoles in the type-IIB picture (sect.~1).

In particular, the N=2 NLSM with the famous (regular and complete) 
{\it Atiyah-Hitchin} (AH) metric \cite{ah} is obtained by the non-conformal 
deformation (6.36) of the N=2 superconformal action (6.33). The corresponding 
PSS data,
$$ \fracmm{1}{2\p i}\oint G (Q_{(4)}(\x),\x) = \fracmm{m}{2\p i} \oint_{C_0} 
\fracmm{Q_{(4)}(\x)}{\x} + \oint_{C_2}\sqrt{Q_{(4)}(\x)}~,\eqno(6.37)$$ 
with the contour $C_0$ encircling the origin in the clock-wise direction, just
 describes the AH metric \cite{ir}. The AH metric also appears in the 
quantum moduli space of the three-dimensional N=4 supersymmetric {\it pure} 
gauge $SU(2)$-based quantum field theory \cite{sw3}, and in the hypermultplet
low-energy effective action (NLSM) of a (magnetically charged) hypermultiplet 
in the Higgs branch \cite{ket}. From the AH prospective, the ALF $D_k$ metrics,
 associated with the HSS potential given by a sum of eqs.~(6.35) and (6.36), 
can be equally interpreted as the deformations of the AH metric. Some of those
metrics were discovered by Dancer \cite{dan}, by using the hyper-K\"ahler 
quotient construction. Their derivation via the generalized Legendre transform
is due to Chalmers \cite{chal} who also noticed the significance of the same
metrics for the monopole moduli spaces in the completely broken $SU(3)$ gauge 
theory investigated by Houghton \cite{hou}.   

To the end of this section, we would like to comment on the geometrical
significance of an $O(4)$ projective multiplet versus an $O(2)$ projective 
(tensor) multiplet, in the context of N=2 supersymmetric NLSM with 
four-dimensional (self-dual or hyper-K\"ahler) target spaces. The PSS 
construction of N=2 NLSM takes the universal form (4.6) for all projective 
multiplets, it generically breaks the $SU(2)_R$ automorphism symmetry of N=2 
supersymmetry algebra, but it leaves a $U(1)$ symmetry. The latter implies an 
$SO(2)$ isometry in the target space of the associated N=2 NLSM. The nature of
 this isometry is, however, dependent upon whether one uses an $O(2)$ or 
$O(4)$ projective multiplet in the PSS action (4.6). Any such action in terms 
of a single $O(2)$ tensor multiplet necessarily leads to a {\it translational} 
(tri-holomorphic) isometry that arises after trading the conserved vector 
component of the $O(2)$ projective multiplet for a scalar by the generalized 
Legendre transform. 

Quite generally, the existence of an isometry amounts to the existence of 
a Killing vector 
$K^m$, $m=1,2,3,4$, obeying eq.~(2.1) or, equivalently, the existence of the 
coordinate system $(x^a,\t)$, $a=1,2,3$, where the metric components are 
independent upon one of the coordinates ($\t$),
$$ ds^2=H^{-1}(d\t +C_a dx^a) + H\g_{ab}dx^adx^b~.\eqno(6.38)$$
A translational isometry implies extra condition on the Killing vector 
\cite{bof},
$$ K^{[m;n]}={}^*K^{[m;n]}~,\eqno(6.39)$$
where the star denotes the four-dimensional dual tensor. Equation (6.39)  
gives rise to the existence of the coordinate system (6.38) where, 
in addition, we have the `monopole equation' \cite{hklr,bof}
$$ \vec{\de}H =\pm \vec{\de}\times\vec{C}~,\quad{\rm and}\quad 
\g_{ab}(x)=\d_{ab}~.\eqno(6.40)$$
Self-duality then amounts to a {\it linear\/} Laplace equation on the 
potential $H(x^a)$,
$$\D H=0~,\eqno(6.41)$$
whose localized solutions,
$$ H(x) =\l + \sum_{s=1}^k\fracmm{m}{\abs{\vec{x}-\vec{x}_s}}~, \eqno(6.42)$$
just describe the $A_k$ series of self-dual metrics (ALE multi-Eguchi-Hanson 
metrics in the case of $\l=0$ and ALF multi-Taub-NUT metrics in the case of 
$\l=1$). It is now not very surprising that those metrics arise from the N=2
superspace NLSM in terms of an $O(2)$ tensor multiplet only. 

However, if one wants to construct the hyper-K\"ahler metrics possessing 
merely a {\it rotational} isometry, the conventional way towards their 
derivation (in components) is much more involved. Equation (6.38) still holds,
 but no eqs.~(6.39), (6.40) and (6.41) are available. Nevertheless, a single 
 real scalar potential for those metrics also exists in  the so-called 
{\it Toda\/} frame defined by the conditions \cite{bof}
$$H=\pa_3\J~,\quad C_1=\pm\pa_2\J~,\quad C_2=\pm\pa_1\J~,\quad C_3=0~,
\eqno(6.43a)$$
and
$$ \g_{11}=\g_{22}=e^{\J}~,\quad \g_{33}=1~.\eqno(6.43b)$$
In the Toda frame self-duality amounts to the {\it non-linear\/} 3d Toda 
equation \cite{bof},~\footnote{The 3d Toda equation arises from the standard 
(2d) $SU(N)$-based Toda system in the large-$N$ \newline ${~~~~~}$ limit 
\cite{savel}.}
$$ \left(\pa^2_1+\pa^2_2\right)\J +\pa^2_3e^{\J}=0~,\eqno(6.44)$$
which is very hard to solve. By the use of an $O(4)$ projective multiplet 
having no conserved vector components, in the N=2 superspace construction of 
4d hyper-K\"ahler metrics, we just deal with the self-dual metrics having 
merely a rotational isometry. The basic geometrical difference between the N=2
 superspace NLSM actions in terms of $O(2)$ or $O(4)$ projective 
multiplets thus amounts to the nature of their abelian isometry: it is 
tri-holomorphic in the $O(2)$ case, whereas it is not triholomorphic in the 
$O(4)$ case. The N=2 NLSM in terms of higher $O(2p)$ projective multiplets are
 similar to that with $p=2$. It is worth mentioning in this context that a
self-dual metric with rotational isometry gives rise to a solution to the 
3d Toda equation (6.44).   

\section{Conclusion}

By the use of harmonic superspace describing hypermultiplets with manifest 
N=2 supersymmetry, we arrived at a general solution to the N=2 non-linear 
sigma-models with special hyper-K\"ahler geometry. Our solution is 
parametrized by a single (of degree two) homogeneous function, which
is quite similar to the well-known N=2 superconformal description of N=2 
vector multiplets in the standard N=2 superspace. 

We also constructed the improved (N=2 superconformal) actions of the $O(2)$
and $O(4)$ projective multiplets, which lead to flat four-dimensional metrics
in disguise. However, after being added together with different moduli, those 
N=2 superconformal actions naturally lead to the $A_k$ and $D_k$ series of the
highly non-trivial self-dual metrics in the target space of the associated N=2
NLSM. It gives us the very natural way of derivation and classification of 
those self-dual metrics. In particular, the ALE metric potentials in 
superspace can be interpreted as the interpolating potentials between 
different improved (flat) potentials in terms of the $O(2)$ projective 
multiplet in the $A_k$ case and in terms of the $O(4)$ projective multiplet in 
the $D_k$ case, whereas the ALF potentials can be understood as non-conformal 
deformations of the ALE ones. It would be interesting to find the self-dual 
metrics, associated with the exceptional (simply-laced) $E_{6,7,8}$ 
Lie groups, in the superspace approach. 

The (improved) N=2 superconformally invariant actions (6.18) and (6.30) of 
the $O(2)$ and $O(4)$ projective multiplets, respectively, are not easily 
generalizable to the case of higher $O(2p)$ projective multiplets with 
$p>2$. An $O(2p)$ projective multiplet is described in HSS by the $L^{(2p+)}$ 
superfield satisfying the off-shell constraints (4.9). It can be used to 
introduce the HSS superfield    
$$ \left[L^{(2p+)}(c^{--})^p\right]^{1/p}  \eqno(7.1)$$
that covariantly transforms under the rigid N=2 superconformal transformations
with conformal weight $w=2$. It is, however, unclear to us how to define a 
covariant HSS superfield of $U(1)$ charge $(+4)$ and conformal weight $w=4$, 
in terms of the $L^{(2p+)}$ superfield, in order to substitute $(L^{++})^2$ 
in eq.~(6.18). This obstruction may be related to the existence of only two 
($A_k$ and $D_k$) regular series of gravitational instantons. It is, however,
 possible to define N=2 superconformal generalizations of eqs.~(6.20) 
and (6.33) to the case of higher projective multiplets, 
$$ S[L^{(2p+)},c]_{\rm ext.} =\int d\z^{(-4)} du (c^{++})^2  
\left[L^{(2p+)}(c^{--})^p\right]^{1/p}~. \eqno(7.2)$$
It would be interesting to study non-conformal deformations of eq.~(7.2).

Our final remark is devoted to {\it local\/} N=2 superconformal symmetry that
implies coupling N=2 (rigidly) superconformal NLSM to N=2 (conformal) 
supergravity. It leads to the deformation of a given special hyper-K\"ahler 
NLSM metric $g_{mn}$ to a quaternionic metric $G_{mn}$ \cite{bagw}. This 
deformation in four dimensions preserves self-duality of the Weyl tensor of the
metric $g_{mn}$,  but it also turns it into an Einstein metric with negative 
scalar curvature \cite{swan}. The associated quaternionic metric reads 
\cite{wkv}
$$ G_{mn}=\fracmm{1}{f}\,g_{mn} -\fracmm{1}{f^2}
\left(\fracm{1}{2}X_mX_n+2Y^A_mY^A_n\right)\eqno(7.3)$$
in terms of the Euler vector $X_m$, the potential $f$ defined by eq.~(2.7), and
the Killing vectors $Y^{(A)}_m$ defined by eq.~(2.9). Therefore, an explicit 
derivation of the quaternionic metric, associated with a given special 
hyper-K\"ahler metric, seems to be possible without going into details of 
the N=2 NLSM coupling to N=2 supergravity.

\section*{Acknowledgements}

It is my pleasure to thank Jan Ambjorn and the High Energy Theory Group at
Niels Bohr Institute for nice hospitality extended to me in Copenhagen 
during the completion of this work. I also thank Luis Alvarez-Gaum\'e, 
Francois Delduc, Olaf Lechtenfeld, Werner Nahm and Galliano Valent for useful 
discussions.


\newpage

\begin{thebibliography}{99}

\bibitem{d3} A. Sen, \np{475}{96}{562};\\
T. Banks, M. Douglas and N. Seiberg, \pl{387}{96}{278};\\
M. Douglas, D. Lowe and J. H. Schwarz, \pl{394}{97}{297}
\bibitem{ads} O. Aharony, S. Gubser, J. Maldacena, H. Ooguri and Y. Oz,
{\it Large N Field Theories, String Theory and Gravity}, hep-th/9905111
\bibitem{af} L. Alvarez-Gaum\'e and D. Z. Freedman, \cmp{80}{81}{443}
\bibitem{hklr} A. Karlhede, U. Lindstr\"om and M. Ro\v{c}ek, 
\pl{147}{84}{297};\\
N.~Hitchin, A.~Karlhede, U.~Lindstr\"om and M.~Ro\v{c}ek,
\cmp{108}{87}{535}
\bibitem{oldk} S. V. Ketov, {\it Selfinteraction of N=2 Multiplets in 4d, 
and UV Finiteness of 2d, N=4 NLSM}, in ``Group-Theoretical Methods in 
Physics'', Moscow: Nauka, 1985, p.~87
\bibitem{gikos} A. S. Galperin, E. A. Ivanov, V. I. Ogievetsky, S. Kalitzin and
E. Sokatchev, \cqg{1}{84}{469} 
\bibitem{wkv} B. de Wit, B. Kleijn and S. Vandoren, {\it Rigid Superconformal
Hypermultiplets}, in ``Supersymmetries and Quantum Symmetries'', edited by
 J. Wess and E. A. Ivanov, Springer, 1999, p.~37, hep-th/9808160;  
 Nucl. Phys. {\bf B568} (2000) 475 
\bibitem{chw} G. Chalmers and A. Hanany, \np{489}{97}{223};\\
 A. Hanany and E. Witten, \np{492}{97}{152}
\bibitem{nahm} W. Nahm, {\it The Construction of All Self-Dual Multi-Monopoles
by the ADHM Method}, in ``Monopoles in Quantum Field Theory'', edited by
N. S. Craigie at al., Singapore: World Scientific, 1982;\\
S. K. Donaldson, \cmp{96}{85}{387};\\ 
J. Hurtubise, \cmp{120}{89}{613};\\
D. Diaconescu, \np{503}{97}{220}
\bibitem{sen} A. Sen, JHEP {\bf 09} (1997) 1; Adv. Theor. Math. Phys. {\bf 1}
(1998) 115
\bibitem{kc} S. A. Cherkis and A. Kapustin, Adv. Theor. Math. Phys. {\bf 2}
(1998) 1287; \cmp{203}{99}{713}
\bibitem{ah} M. Atiyah and N. Hitchin, The Geometry and Dynamics of Magnetic
Monopoles, Princeton University Press, 1988
\bibitem{kron} P. B. Kronheimer, J. Diff. Geometry {\bf 29} (1989) 665; 685
\bibitem{fer} A. Ferber, \np{132}{78}{55}
\bibitem{giosc} A. S. Galperin, E. A. Ivanov, V. I. Ogievetsky and 
E. Sokatchev, {\it Conformal Invariance in Harmonic Superspace}, in ``Quantum
Field Theory and Quantum Statistics'', edited by C. J. Isham and G. Vilkovisky,
Adam Hilger, 1987, Vol.~2, p.~233   
\bibitem{gr} G. W. Gibbons and P. Rychenkova, \pl{443}{98}{138}
\bibitem{kw} I. Klebanov and E. Witten, \np{356}{98}{199}
\bibitem{blair} D. Blair, Compact Manifolds in Riemannian Geometry,
Springer, 1976
\bibitem{bg} C. P. Boyer and K. Galicki, {\it 3-Sasakian Manifolds}, to appear
in ``Essays on Einstein Manifolds'', edited by M. Wang et al.; hep-th/9810250
\bibitem{afq} B. Acharya, J. Figueroa-O'Farrill, C. Hull and B. Spence,
Adv. Theor. Math. Phys. {\bf 2} (1999) 1249
\bibitem{rec} S. V. Ketov, {\it Exact Low-Energy Effective Actions for
Hypermultiplets in Four Dimensions}, hep-th/9909177; to appear in 
Int. J. Mod. Phys. {\bf A}.   
\bibitem{giosk} A. S. Galperin, E. A. Ivanov, V. I. Ogievetsky and 
E. Sokatchev, Ann. Phys. (N.Y.) {\bf 185} (1988) 1; 22
\bibitem{met}  A. S. Galperin, E. A. Ivanov, V. I. Ogievetsky and 
E. Sokatchev, \cmp{103}{86}{515};\\
A. S. Galperin, E. A. Ivanov, V. I. Ogievetsky and  P. K. Townsend, 
\cqg{3}{86}{625};\\
G. W. Gibbons, D. Olivier, P. J. Ruback and G. Valent, \np{296}{88}{679};\\
S. V. Ketov and Ch. Unkmeir, \pl{422}{98}{179}
\bibitem{fan} S. J. Gates Jr., \np{238}{84}{349};\\
S. V. Ketov and I. V. Tyutin, JETP Lett. {\bf 39} (1984) 703 
\bibitem{van} B. de Wit, J. W. van Holten and A. van Proeyen, 
\np{167}{80}{186};\\
B. de Wit, P. Lauwers and A. van Proeyen, \np{255}{85}{569}
\bibitem{ten} A. S. Galperin, E. A. Ivanov and V. I. Ogievetsky, 
\np{282}{87}{74}  
\bibitem{imc} B. de Wit, R. Philippe and A. van Proeyen, \np{219}{83}{143}
\bibitem{imp} U. Lindstr\"om and M. Ro\v{c}ek, \np{222}{83}{285}
\bibitem{imh} A. S. Galperin, E. A. Ivanov and V. I. Ogievetsky, Sov. J. Nucl.
Phys. {\bf 45} (1987) 157
\bibitem{ir} I. T. Ivanov and M. Ro\v{c}ek, \cmp{182}{96}{291}
\bibitem{crw} G. Chalmers,  M. Ro\v{c}ek and S. Wiles, JHEP {\bf 01} (1999) 009
\bibitem{sw3} N. Seiberg and E. Witten, {\it Gauge Dynamics and 
Compactification to Three Dimensions}, in ``Saclay'96. The Mathematical Beauty
of Physics''; hep-th/9607163
\bibitem{ket} S. V. Ketov, \pl{469}{99}{136}
\bibitem{dan} A. S. Dancer, Quart. J. Math. Oxford. {\bf 45} (1994) 463
\bibitem{chal} G. Chalmers, Phys. Rev. {\bf D58} (1998) 125011
\bibitem{hou} C. J. Houghton, Phys.Rev. {\bf D56} (1997) 1220.
\bibitem{bof} C. Boyer and J. Finley, J. Math. Phys. {\bf 23} (1982) 1126;\\
J. Gegenberg and A. Das, Gen. Rel. Grav. {\bf 16} (1984) 817;\\
I. Bakas and K. Sfetsos, Int. J. Mod. Phys. {\bf A12} (1997) 2585
\bibitem{savel} M. Saveliev, \cmp{121}{89}{283}; Theor. Math. Phys. {\bf 92}
 (1993) 1024
\bibitem{bagw} J. Bagger and E. Witten, \np{222}{83}{1}
\bibitem{swan} A. Swan, Math. Ann. {\bf 289} (1991) 421;\\
K. Galicki, Class. Quantum Grav. {\bf 9} (1992) 27.
\end{thebibliography}
\end{document}